# An Overview of the Decimation process and its VLSI implementation

Rozita Teymourzadeh  & Masuri Othman
Department of Electrical, Electronic and Systems Engineering
VLSI Design Research Group
National University of Malaysia
rozita60@vlsi.eng.ukm.my

## ABSTRACT

Digital Decimation process plays an important task in communication system. It mostly is applied in transceiver when the frequency reduction is required. However, the decimation process for sigma delta modulator is considered in this research work. The proposed design was simulated using MATLAB software  and implemented by hardware description language in Xilinx environment.  Furthermore, the proposed advance arithmetic unit is applied to improve the system efficiency.

## INTRODUCTION

Although real world signals are analog, but digital to analog converter (ADC) helps lead signal to digital domain due to it is easier evaluate and process. Digital signal can be converted back to analog signal by digital to analog converter (DAC). Over sampling modulator is applied for audio application to convert it as digital signal by high sampling frequency. The audio signal is sampled within the modulator at a rate significantly higher than the Nyquist rate. After over sampling, decimation block is required to remove noise shaping and decimate the digital signal from high to low. For over sampling, Sigma delta modulator and for decimation Cascaded Integrator Comb filter (CIC), half band and Finite Impulse Response filter have been selected to carry out the task. This paper describes decimation process by the focus on high speed implementation of CIC filter (Hogenauer EB 1981). The advantages of this paper are using three methods to speed up the CIC filter such as using Modified Carry Look-adder Adders and achieve the filter ripple less than 0.0002 db. Simulink toolbox available in Matlab software which is used to simulator and Verilog HDL coding by Xilinx software help to verify the functionality of the CIC filters and Implement it on VLSI as chip.

## DESCRIPTION

Digital audio application such HiFi CD and DAT systems often use sigma delta A/D converters (Aziz,, Sorensen & Spiegel 1996). The quality of sigma delta modulator is recognized by the order and its resolution. In this project the sampling frequency represent as 6.144 MHz with the over sampling ration of 128. Nyquist frequency is selected to be 48 kHz to support $f_B$ equal to 24 kHz with the frequency response ripple less than

0.0002 db. Figure 1 shows 3rd order sigma delta modulator with multirate decimation filter. A multirate decimation filter system was chosen to realize the needed performance. The filter system was thus organized as an initial filter stage having a 16:1 decimation ratio followed by a third stage having an 8:1 decimation ratio.

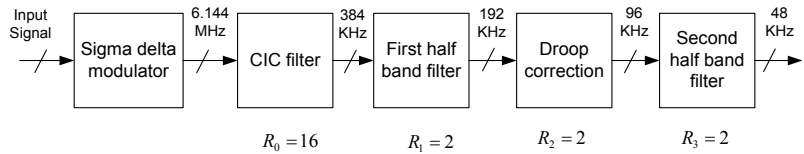

$R_0 = 16 \quad R_1 = 2 \quad R_2 = 2 \quad R_3 = 2$

FIGURE 1: Digital Decimation Process

CIC filter is located after sigma delta modulator and decimate the frequency by the ratio of 16. The packing of modulator and CIC filter minimized the noise by decrease the number of parallel pad drivers. Then CIC filter increase the sigma delta resolution to improve Signal to Noise ratio. The two half band filters (Brandt & Wooley 1994) are used to reduce remain sampling rate reduction to the Nyquist output rate. First half band filter and second half band filter make the frequency response more flat and sharp similar to ideal filter specially second half band filter due to higher order of the filter (R=40), has most effect to make the frequency response sharp. All the even coefficients of half band filters are zero exception last one which is 0.5. This particular make them efficient for 2:1 decimation ratio and reduce the computational complexity by near 50% as compared to general direct form filter Architecture. Droop correction filter is allocated to compensate pass band attenuation which is created by CIC filter. The frequency response of overall system will be shown in Figure 2.

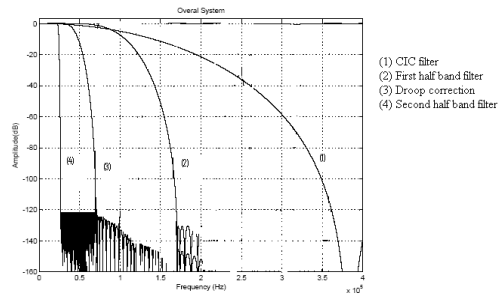

FIGURE 2: Frequency response of decimation system

Similar Decimation process has been done by Thompson (Thompson 1989) but in this paper high speed Cascaded Integrator Comb filter is designed and implemented to accomplish decimation task, remove quantization noise and avoid aliasing to the signal.

## CASCADED INTEGRATOR COMB FILTER

Previously, Low pass decimation filter structure (FIGURE 3) widely used for decimation before appearance of the CIC decimation filter (Adams 1994). This structure has M Multipliers which make the large number of computation whereas new structure of decimation or CIC is multiplier less in terms of minimizing hardware and computational. Additionally the CIC filter does not require storage for filter coefficients and multipliers as all coefficients are unity (Sangil Park 1990). The VLSI implementation of CIC filter makes it able to be used as a SoC chip.

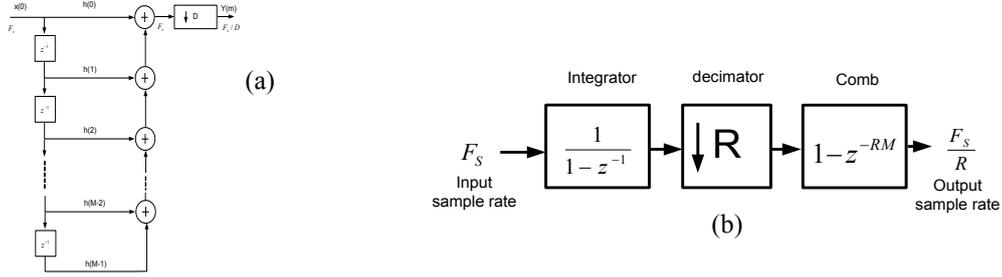

FIGURE 3: Decimation filter (a) low pass decimation filter (b) CIC filter where N is the number of stage, M is the differential delay and R is the decimation factor

The CIC filter consist of N stages of integrator and comb filter which are connected by a down sampler stage as shown in figure 1 in z domain. The CIC filter has the following transfer function:

$$H(z) = H_I^N(z).H_C^N(z) = \frac{(1-z^{-RM})^N}{(1-z^{-1})^N} = (\sum_{k=0}^{RM-1} z^{-k})^N \quad (1)$$

In this paper, N, M and R have been chosen to be 5, 1 and 16 respectively to avoid overflow in each stages. CIC filter has low and high pass component. Integrator part with low pas transfer function structure amplify low frequency component so integrator it self is not stable due to the integrator output will cause over flow. In this case, the comb stage with high pass structure attenuates the low frequency component, making the whole system stable. N, M and R are parameters to determine the register length requirements necessary to assure no data loss. The maximum register growth/width, $G_{max}$ can be expressed as:

$$G_{max} = (RM)^N \quad (3)$$

In other word, $G_{max}$ is the maximum register growth and a function of the maximum output magnitude due to the worst possible input conditions (Hogenauer EB 1981).

If the input data word length is $B_{in}$, most significant bit (MSB) at the filter output, $B_{max}$ is given by:

$$B_{max} = [N\log_2 R + B_{in} - 1] \quad (4)$$

In order to reduce the data loss, normally the first stage of the CIC filter has maximum number of bit compared to the other stages.

## NOVEL SPECIFICATIONS

This paper describes how to enhance the decimation system. To achieve the aim, three methods are used as follow:

### 1. High speed CIC filter

#### 1.1 Truncation

Truncation means estimating and removing Least Significant Bit (LSB) to reduce the area requirements on chip and power consumption and also increase speed of calculation. Although this estimation and removing introduces additional error, the error can be made small enough to be acceptable for Audio applications. Figure 2 illustrates five stages of the CIC filter when $B_{max}$ is 25 bit so truncation is applied to reduce register width. Matlab software helps to find word length in integrator and comb section.

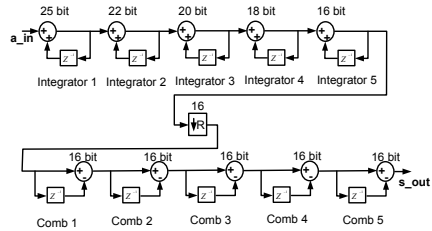

FIGURE 4: Five-stages of truncated CIC filter include integrator and comb cell

## 1.2 Modified Carry look-ahead adder (MCLA)

The other technique to increase speed is using Modified Carry Look-ahead Adder (Ciletti 2003). The Carry Look-ahead adder (CLA) is the fastest adder which can be used for speeding up purpose but the disadvantage of the CLA adder is that the carry logic is getting quite complicated for more than 4 bits so Modified Carry Look-ahead Adder (MCLA) is introduced to replace as adder. This improve in speed is due to the carry calculation in MCLA. The 25bit MCLA structure is shown in Figure 4. Its block diagram consists of 2, 4-bit module which is connected and each previous 4 bit calculates carry out for the next carry. The Verilog code has been written to implement summation. The MCLA Verilog code was downloaded to the Xilinx FPGA chip. It was found minimum clock period on FPGA board is 4.389ns (Maximum Frequency is 220 MHz).

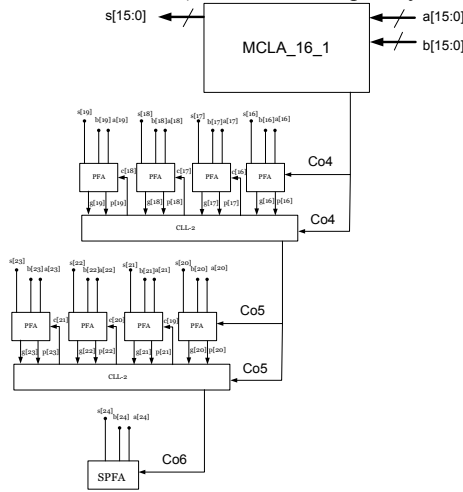

FIGURE 5: The 25 bit MCLA structure

## 1.3 Pipeline structure

One way to have high speed CIC filter is by implementing the pipeline filter structure. Figure 6 shows pipeline CIC filter structure. In the pipelined structure, no additional pipeline registers are used for integrator part. (Djadi et al 1994).

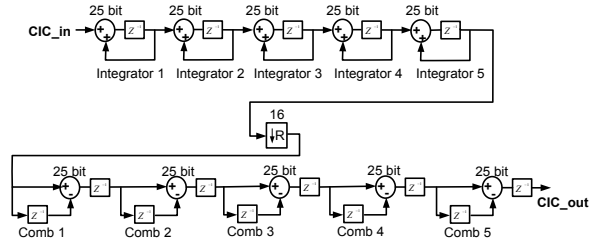
FIGURE 6: Five-stage of truncated pipeline CIC filter include integrator and comb cell

## 2. FILTER RIPPLE

Ripple is usually specified as a peak-to-peak level in decibels. It describes how little or how much the filter's amplitude varies within a band. Smaller amounts of ripple represent more consistent response and are generally preferable. There is, however, a tradeoff between ripple and transition bandwidth, so that decreasing either will only serves to increase the other. The way of ripple improvement is increasing the number of coefficients used by the filter. One disadvantage of using increasingly long filter lengths is the compute time required to perform the filtering. In this paper the order of first half band filter, droop correction and second half band filter has been design to be 4,8 and 40 respectively to provide ripple less than 0.0002 db.

## DESIGN RESULTS

Figure 7 shows the Droop correction filter result. This filter design a low pass filter with pass band having the shape of inverse CIC filter frequency response. So it compensates amplitude droop cause of the CIC filter and makes whole system frequency response flat.

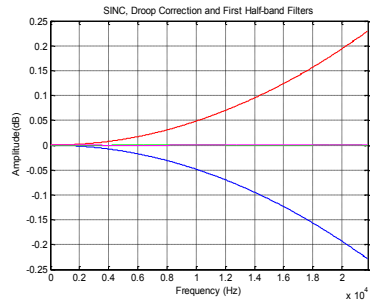
FIGURE 7: Droop Correction effect on frequency response

Figure 8 shows the measured baseband output spectra before (Figure 8(a)) and after (Figure 8(b)) the decimation functions. The CIC filter Verilog code was wrote and simulated by Matlab software. It is found Signal to Noise ratio (SNR) is 141.56 dB in sigma delta modulator output and SNR is increased to 145.35 dB in the decimation stages. To improve the signal to noise ratio, word length of recursive CIC filter should be increased but the speed of filter calculation is also decreased.

FIGURE 8: Signal spectra (a) Output $\sum\Delta$ modulator SNR (b) Output CIC filter SNR

Figure 9 shows the pass band ripple for whole decimation system. It is clear that the pass band ripple is less than 0.0002 db.

FIGURE 9: Pass band ripple in whole decimation frequency response

FIGURE 10: Simulation result on FPGA board

The speed of CIC filter is improved to 163 MHz compared to behavioural CIC filter which is 107 MHz. the Design Analyzer software under Synopsis shows that the comb stage make maximum delay in decimation process.

## CONCLUSIONS

Recursive CIC filters have been designed and investigated. Enhanced high Speed CIC filters was obtained by truncation, the pipeline structure and by using the modified carry look-ahead adder (MCLA). The evaluation indicates that the pipelined CIC filter with MCLA adder is attractive due to high speed when both the decimation ratio and filter order are not high as stated in the Hogenauer Comb filter.

## ACKNOWLEDGEMENT

I wish to thank my supervisor, Prof. Dr. Masuri Othman who has assisted me for completion of this dissertation.

## REFERENCES

Hogenauer EB, (1981). *An economical class of digital filters for decimation and interpolation,* IEEE transactions on acoustic, Sunnyvale, CA.Assp-29(2):155162

Pervez M. Aziz, Henrik V. Sorensen & Jan Van Der Spiegel, (1996) *An Overview of Sigma –Delta Converter,* IEEE Signal processing magazine, 1053-5888/96, 61-82

Charles D. Thompson, (1989). *A VLSI Sigma Delta A/D Converter for Audio and Signal Processing Applications,* IEEE, Motorola DSP Operations, Austin, Texas, CH2673-2/89/0000-2569 IEEE.

Brian P. Brandt and Bruce A. Wooley, (1994) *A Low-Power, Area-Efficient Digital Filter for Decimation and Interpolation*, IEEE Journal of Solid-State Circuits, Vol. 29, No.6

R.Adams, (1994). *Design aspects of high-order delta-sigma A/D converters*, IEEE International Symposium on Circuits and Systems Tutorials, pp. 235-259.

Sangil Park, (1990). *Principles of Sigma-delta Modulation for Analog-to-Digital Converters*, Motorola Inc, APR8/D Rev.1.

Michael D. Ciletti (2003), *Advanced Digital design with the Verilog HDL*, Prentice Hall, Department of Electrical and Computer Engineering University of Colorado at Colorado Springs

Y. Djadi and T. A. Kwasniewski, C. Chan and V. Szwarc, (1994). *A high throughput Programmable Decimation and Interpolation Filter.* Proceeding of International Conference on Signal Processing Applications and Technology, pp.1743-1748.